\documentclass[conference]{IEEEtran}
\IEEEoverridecommandlockouts

\usepackage{cite}
\usepackage{amsmath,amssymb,amsfonts}
\usepackage{algorithmic}
\usepackage{graphicx}
\usepackage{textcomp}
\usepackage{xcolor}
\def\BibTeX{{\rm B\kern-.05em{\sc i\kern-.025em b}\kern-.08em
    T\kern-.1667em\lower.7ex\hbox{E}\kern-.125emX}}
\usepackage[caption=false,font=footnotesize]{subfig}
\usepackage{booktabs}
\usepackage{multirow}
\usepackage{makecell}
\begin{document}

\title{Investigating the Impact of Speech Enhancement on Audio Deepfake Detection in Noisy Environments\\
}



\author{%
\makebox[\textwidth][c]{%
\begin{tabular}{@{}p{0.32\textwidth}@{\hspace{0.02\textwidth}}p{0.32\textwidth}@{\hspace{0.02\textwidth}}p{0.32\textwidth}@{}}

\centering
{\large Anacin, Angela}\\
\textit{School of Computing}\\
Wichita State University\\
Wichita, Kansas, USA\\
aranacin@shockers.wichita.edu
&
\centering
{\large Shruti Kshirsagar}\\
\textit{School of Computing}\\
Wichita State University\\
Wichita, Kansas, USA\\
shruti.kshirsagar@wichita.edu
&
\centering
{\large Anderson R. Avila}\\
Institut national de la recherche scientifique (INRS--EMT)\\
Montreal, Canada\\
anderson.avila@inrs.ca

\end{tabular}%
}%
}
\maketitle
\thispagestyle{empty}
\pagestyle{empty}


\begin{abstract}
Logical Access (LA) attacks, also known as audio deepfake attacks, use Text-to-Speech (TTS) or Voice Conversion (VC) methods to generate spoofed speech data. This can represent a serious threat to Automatic Speaker Verification (ASV) systems, as intruders can use such attacks to bypass voice biometric security. In this study, we investigate the correlation between speech quality and the performance of audio spoofing detection systems (i.e., LA task). For that, the performance of two enhancement algorithms is evaluated based on two perceptual speech quality measures, namely Perceptual Evaluation of Speech Quality (PESQ) and Speech‑to‑Reverberation Modulation Ratio (SRMR), and in respect to their impact on a state-of-the-art audio spoofing detection system, known as Audio Anti-Spoofing using Integrated Spectro-Temporal Graph Attention Networks (AASIST). We adopted the LA dataset, provided in the ASVspoof 2019 Challenge, and corrupted its test set with different Signal-to-Noise Ratio (SNR) levels, while leaving the training data untouched. Enhancement was applied to attenuate the detrimental effects of noisy speech, and the performances of two models, Speech Enhancement Generative Adversarial Network (SEGAN) and Metric-Optimized Generative Adversarial Network Plus (MetricGAN+), were compared. Although we expect that speech quality will correlate well with speech applications' performance, it can also have as a side effect on downstream tasks if unwanted artifacts are introduced or relevant information is removed from the speech signal. Our results corroborate with this hypothesis, as we found that the enhancement algorithm leading to the highest speech quality scores, MetricGAN+, provided the lowest Equal Error Rate (EER) on the audio spoofing detection task, whereas the enhancement method with the lowest speech quality scores, SEGAN, led to the lowest EER, thus leading to better performance on the LA task. 
\end{abstract}

\begin{IEEEkeywords}
Audio Deepfake Detection, Speech Enhancement, Logical Access, Noise-Robustness.
\end{IEEEkeywords}

\section{Introduction}
\label{sec:intro}

User authentication based on voice biometrics has already become a practical reality. Embedded with Automatic Speaker Verification (ASV) capabilities, voice-enabled devices such as Google Home and Amazon Alexa can authenticate users through their voices \cite{arif2021voice}. These systems are also being used by financial institutions as a security layer in multi-factor authentication (MFA), enhancing security while ensuring user convenience \cite{oyewale2024optimizing, narang2024artificial}. In fact, it is found that ASV offers a more natural way of authentication, capitalizing on our voice, which is considered the most preferable biometric modality among consumers \cite{avila2021automatic}. Nevertheless, these systems remain susceptible to adversarial attacks designed to manipulate speech data to bypass biometric security systems. 

In response to these vulnerabilities,  a series of benchmarking initiatives have been made to support the development of countermeasures against ASV spoofing attacks \cite{wu2017asvspoof}. Since 2015, many efforts to create more realistic scenarios to support the development of robust countermeasure systems have been made. For example, ASVspoof 2021 simulated the transmission of bonafide and spoofed speech across telephony systems, including a voice-over-internet-protocol (VoIP) and a public switched telephone network (PSTN) \cite{yamagishi2021asvspoof}\cite{liu2023asvspoof}. Additive and convolutive noises, which are typically encountered in our everyday surroundings, remained left out. To the best of our knowledge, the robustness of synthetic speech detection under acoustically degraded conditions is still overlooked in the literature, and only a limited number of studies have addressed this issue. In 
\cite{hanilci2016spoofing}, for instance, the authors have conducted extensive experiments to assess the robustness of existing state-of-the-art countermeasure systems. They used the ASVspoof 2015 data and concluded that traditional enhancement systems, such as magnitude spectral subtraction, power spectral subtraction, and Wiener filtering, offered limited contribution towards improving the accuracy of detection systems. Researchers have been able to introduce a novel dataset, termed Audio Deepfake Detection (ADD) \cite{yi2022add}, which incorporates diverse background noise into deepfake audio. In a more recent work \cite{chen2020generalization}, additive and convolutive noises were considered as a data augmentation strategy and not as impairment during the system evaluation. Results are reported on the ASVspoof 2019 dataset and improvements are achieved with three main strategies: (1) Frequency masking where consecutive frequency bands are dropped during training; (2) large margin cosine loss to enforce the model to learn embeddings that maximize the inter-class variance while minimizing the intra-class variance and (3) data augmentation by using background noise and reverberation during training. A similar work is reported in \cite{das2021data}, where a new data augmentation technique based on a-law and mu-law was proposed for the Logical Access (LA) task. Similarly, in \cite{tak2022rawboost}, a data augmentation based on the combination of linear and non-linear convolutive noise, impulsive signal-dependent additive noise and stationary signal-independent additive noise, is proposed to improve the performance of systems designed for the LA task.

It is important to note that these solutions fall short in addressing the detrimental effects of background noise or the impact of enhancement on the LA task. Our research seeks to bridge this gap by assessing not only the impact of background noise but also the influence of enhancement techniques commonly used in speech communication on the LA task. Furthermore, we examined how speech quality correlates with the performance of a state-of-the-art spoofing detection system, namely Audio Anti-Spoofing using Integrated Spectro-Temporal Graph Attention Networks (AASIST) \cite{jung2022aasist}, adopted as a  benchmark in this work. For that, we adopted the data provided by the ASVspoof 2019 edition, which contains only clean speech. That is, all signals are free from additive noise, reverberation and network impairments \cite{liu2023asvspoof}. This is not realistic as during acquisition, transmission, and reception, digital signals are typically affected by the acoustic environment and communication channels. To simulate more realistic scenarios (i.e., beyond laboratory settings), we evaluate the performance of the adopted countermeasure benchmark system under noisy conditions. For that, the test set of the ASVspoof 2019 dataset is corrupted with background noise. Similar to the work proposed in \cite{avila2018investigating, kshirsagartask}, we explore two enhancement algorithms based on a deep neural network to mitigate the detrimental effects of background noise. We also report the correlations between the estimated speech perceptual quality measures and the performance of the countermeasure model under different Signal-to-Noise Ratio (SNR) levels.  

The remainder of this paper is organized as follows. Section \ref{sec:new_template} presents materials and methods used in this study. In Section \ref{sec:Ex_ds}, we present the experimental setup and in Section \ref{sec:results} we present our results and discussion. Section \ref{sec:conclusion} concludes the article.
\section{Methods and Materials}
\label{sec:new_template}

Speech communication systems are typically equipped with speech enhancement meant to eliminate unwanted artifacts, such as background noise and reverberation. Here, two noise suppression algorithms are presented along with two speech quality measures. We also give details on the spoofing countermeasure model adopted for this study.

\subsection{Speech Enhancement Methods}
\label{subsubsec:metricgan}

The first enhancement method investigated is the Metric-Optimized Generative Adversarial Network Plus (MetricGAN+) \cite{fu2021metricgan+}. Following its predecessor, namely Metric-Optimized Generative Adversarial Network (MetricGAN) \cite{fu2019metricgan}, the authors proposed three main adjustments on the previous architecture to attain better speech quality of enhanced signals. The first adjustment was the addition of noisy speech during the optimization of the discriminator. They also proposed a replay buffer, which referred to the use of generated samples as training data. The authors proposed optimizing the sigmoid mapping used in the loss function to improve learning during training. MetricGAN+ was shown to generate speech samples that led to higher perceptual speech quality scores compared to other methods, specifically its predecessor, MetricGAN. The second enhancement method investigated here is the Speech Enhancement Generative Adversarial Network (SEGAN) \cite{pascual2017segan}. It was first introduced as one of the first frameworks that used Generative Adversarial Networks (GANs) to improve speech. Its purpose is to improve speech quality by reducing noise in corrupted speech signals, with an emphasis on enhancing both intelligibility and perceptual quality. The system employs an encoder-decoder model with skip connections to reconstruct clean speech from noisy input signals.

\subsection{Instrumental Quality Measures}

Two speech quality measures are considered in this work. An intrusive one, namely Perceptual Evaluation of Speech Quality (PESQ) \cite{rix2001perceptual}, and a non-intrusive one, named Speech‑to‑Reverberation Modulation Ratio (SRMR) \cite{santos2014improved}). PESQ is consider an intrusive quality measure, as it requires the reference signal to estimate the speech quality. The pipeline of the PESQ algorithm begins by aligning the reference and processed signals with respect to their magnitude level and time. The level-aligned signals are then passed through a perceptually motivated auditory transformation. Disturbances are aggregated in frequency and time and then assigned to the MOS scale \cite{de2024pesqetarian}. The second speech quality measure adopted is the SRMR. It computes the ratio of low to high modulation energy after an auditory model is applied using a gammatone filterbank that emulates the human hearing system. The measure relies on the principle that the modulation energy of clean speech is generally concentrated in lower modulation frequencies (below 20 Hz) while room acoustic artifacts typically arise in higher modulation frequencies beyond 20 Hz. Although tailored for reverberation, the model has been shown to accurately characterize the quality of speech processed by enhancement algorithms.

\subsection{Spoofed Speech Detection Model}
\label{subsec:Lightweight Model}
We adopt the AASIST for the LA task. The AASIST model leverages Graph Attention Networks (GATs) and is an enhanced version of RawGAT-ST, the state-of-the-art model from the ASVspoof 2019 Challenge. It surpasses previous models in spoofing detection accuracy. Central to AASIST's architecture is a heterogeneous stacking graph attention layer, designed to model artifacts across temporal and spectral domains using a novel heterogeneous attention mechanism and stack node. Additionally, the model incorporates an innovative max graph operation that employs a competitive mechanism alongside an extended readout scheme. A RawNet2-based encoder is used to encode 64,600 raw waveforms (about 4 seconds) during implementation. This encoder starts with a sinc-convolution layer with 70 filters, followed by six residual blocks, two with 32 filters and four with 64 filters.
The graph attention layers are designed for the  best performance. The first two layers use 64 filters, whereas graph-pooling layers cut spectral and temporal nodes by 50\% and 30\%, respectively. Graph attention layers with 32 filters are followed by graph-pooling processes, which reduce nodes by 50\%. \cite{jung2022aasist}

\section{Experimental Details}
\label{sec:Ex_ds}

\subsection{Corpus Description}
In our experiments, we use the Logistic Access (LA) subset of the ASVspoof 2019 database, which was built with the effort of using a  diverse pool of spoofing algorithms. It includes 17 different  Text-to-Speec (TTS) and Voice Conversion (VC) systems. Six of these systems (A01-A06) are known attacks, whereas the other eleven (A07-A19, excluding A16 and A19) are unknown. These systems use a number of approaches, including neural network-based TTS and VC systems, waveform concatenation techniques, and hybrid methods. Some systems use complex neural vocoders like WaveNet and WORLD, whilst others use simpler or faster options like Griffin-Lim or Vocaine. Known systems, like as A01 and A02, use powerful neural frameworks and vocoders to generate high-quality speech that tests countermeasures. The unknown systems build upon these approaches by including techniques such as GAN-based post-filters (A07), end-to-end TTS systems (A10), and cutting-edge VC frameworks (A17). A10 and A11 use Tacotron 2 for TTS, with differences in vocoder technology, whilst A17 uses direct waveform modification for improved spoofing capability. Many systems utilize transfer learning, such as A10's speaker embeddings based on ASV tasks. Other breakthroughs include non-parallel VC approaches (A05 and A18), which generate transformed speech without the need for paired data by leveraging latent vectors or i-vector frameworks. These systems demonstrate the progression of spoofing technologies, from traditional waveform concatenation (A04 and A16) to complex neural architectures that improve naturalness and spoofing efficacy. The diversity underlines the difficulty in building effective responses, as attackers increasingly use tactics that resemble legitimate speech characteristics. This trend emphasizes the need for continued study into identifying and countering such advanced spoofing attacks.

\subsection{Benchmark and Evaluation Setup}
\label{se:setup}
\begin{figure}
\centerline{\includegraphics[width=0.48\textwidth]{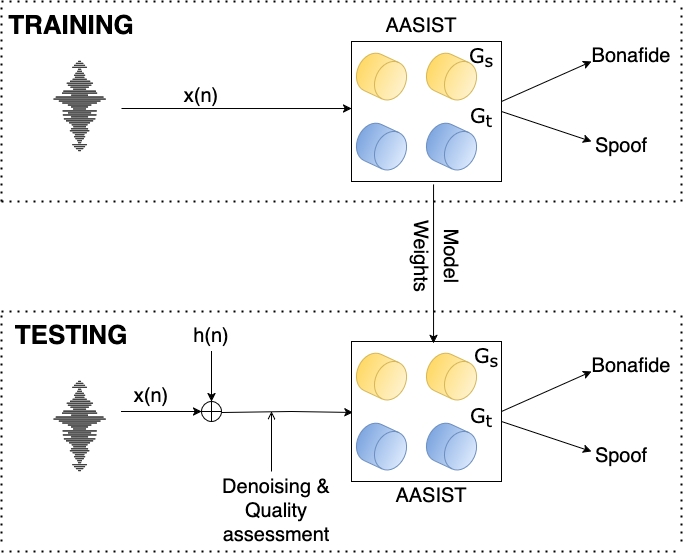}}
\caption{Training and Testing Setup; x(n) refers to the raw audio, h(n) refers to convolutive noise being added, Gs and Gt refer to Graph Spectral and Graph Temporal, these combine and pass into the Heterogeneous Stacking Graph Attention Layer (HS-GAL).}
\label{setup}
\end{figure}

As seen in Figure \ref{setup},  our training configuration involves input of raw audio to the AASIST model, which is used as our benchmark detection model. It subsequently provides the results of spoofing detection. Subsequently, we employ the trained model to evaluate unfamiliar noisy samples. Noise was added only in the testing setup to show the effect of noise. In addition, we employ enhancements to eliminate some noise from the data and assess the audio quality to preserve essential elements inside the audio. The enhanced noise is subsequently processed via the identical pretrained model to assess the impact of enhancement on noisy audio for spoofing detection.

\subsection{\textbf{Corrupted Speech and Enhancement}}
\label{se:noisy}
\begin{figure}
\centerline{\includegraphics[width=0.48\textwidth]{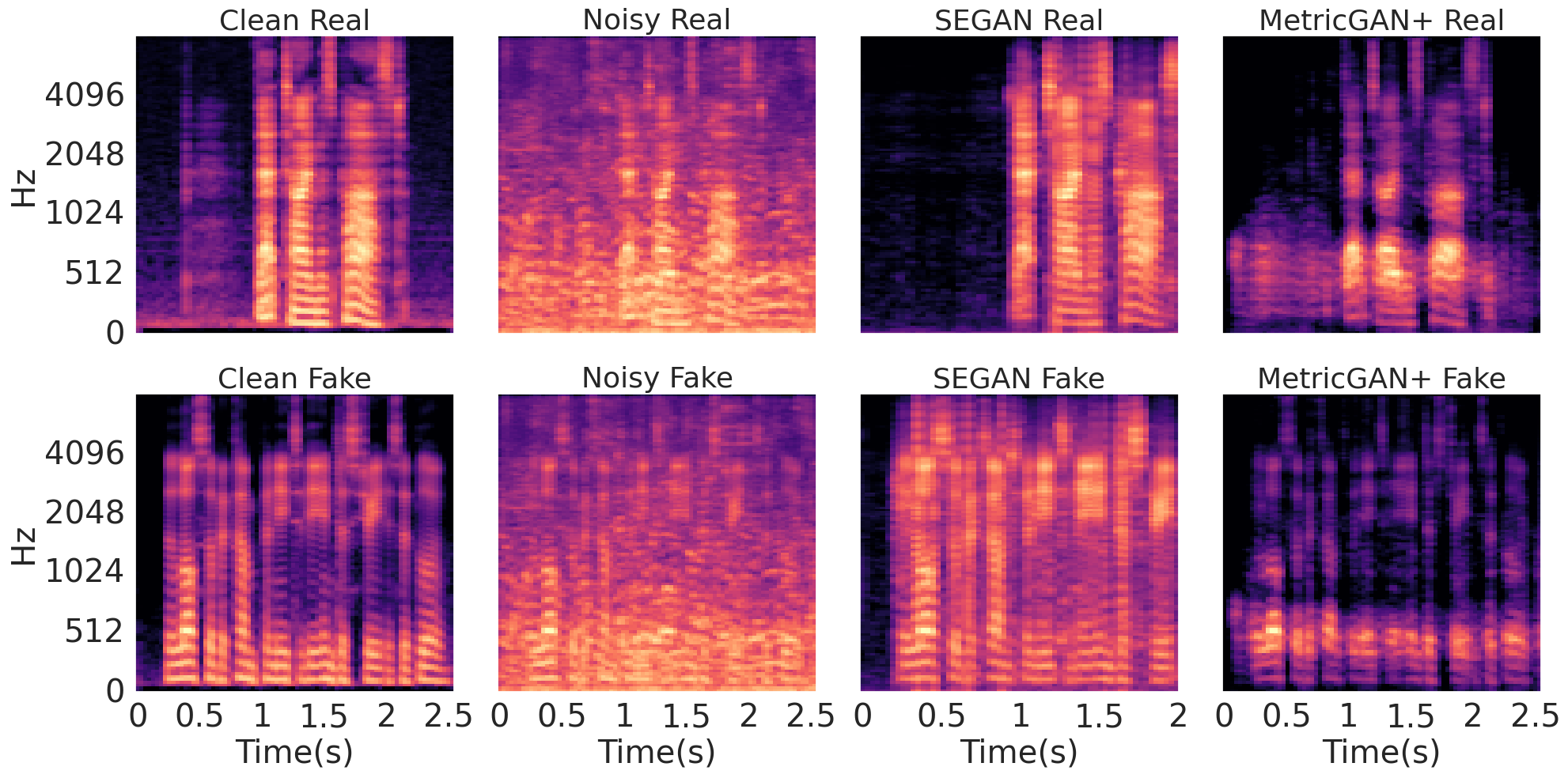}}
\caption{SNR for real and spoof audio signal at 0dB}
\label{noisySNR}
\end{figure}

To evaluate noise robustness, we created an artificially generated noisy corpus from the ASVspoof 2019 challenge. We used two noisy environments: Babble and Cafeteria.  Each additive noise type has five different SNR levels: 0 dB, 5 dB, 10 dB, 15 dB, and 20 dB. Figure \ref{noisySNR} shows the SNR for Cafeteria and Babble at 0dB, highlighting the impact of noise. The spectrogram clearly indicates that noise has completely overrun both the clean and spoof audio signals.

We tested enhanced noise robustness using artificially generated noisy data from the ASVspoof 2019 competition, as previously reported. To support this, we used SEGAN and MetricGAN+. SEGAN was chosen to compare the results with MetricGAN+, despite the fact that MetricGAN+ provides a value comparable to human perception or PESQ/ Short-time objective intelligibility (STOI) ratings for references, but SEGAN evaluates upgrades using a min-max loss technique. Figure \ref{noisySNR} displays the SNR for enhanced Cafeteria and Babble at 0dB, demonstrating the influence of enhancement. The spectrogram clearly shows that improvements may reveal the clean signal considerably better for both clean and spoof audio signals. SEGAN used pretrained weights for enhancement, while MetricGAN+ was sourced from the SpeechBrain toolkit \cite{speechbrain}. Noisy audio was processed using the pretrained models, resulting in enhanced versions that were subsequently input into the AASIST model. 

\subsection{Figures-of-merit}
\label{se:metrics}
To assess the performance of the benchmark model, Equal Error Rate (EER) and tandem detection cost function (t-DCF) have been used. EER relies on false negative rate (FNR) and false positive rate (FPR) and a threshold \cite{todisco2019asvspoof}. The measure represents that the proportion of false acceptances is equal to the proportion of false rejections. The  t-DCF, on the other hand, evaluates the performance of a cascaded system that includes a Countermeasure (CM) and an ASV system \cite{todisco2019asvspoof}. It calculates the cost of errors using prior probability for speaker categories (target, non-target, spoof), the system's error rates, and user-defined cost parameters for missed detections and false alarms. 

\begin{figure}[!t]
    \centering
    \subfloat[Cafeteria]{%
        \includegraphics[width=0.48\columnwidth]{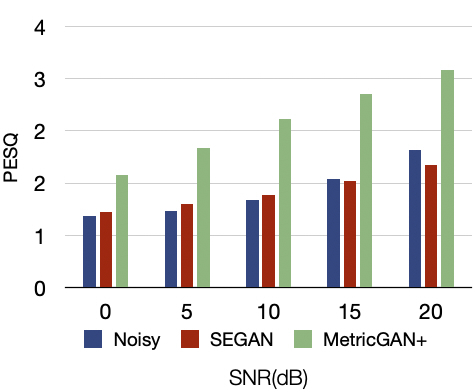}%
        \label{fig:cafeteria1}}
    \hfil
    \subfloat[Babble]{%
        \includegraphics[width=0.48\columnwidth]{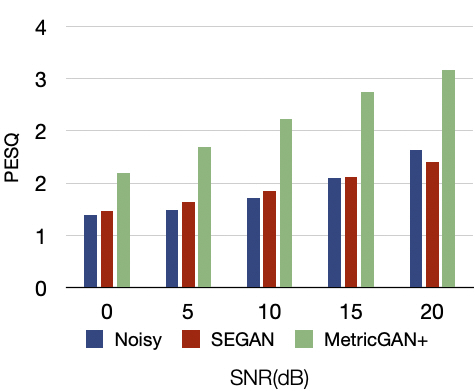}%
        \label{fig:babble1}}
    \\[1ex] 
    \subfloat[Cafeteria (SRMR)]{%
        \includegraphics[width=0.48\columnwidth]{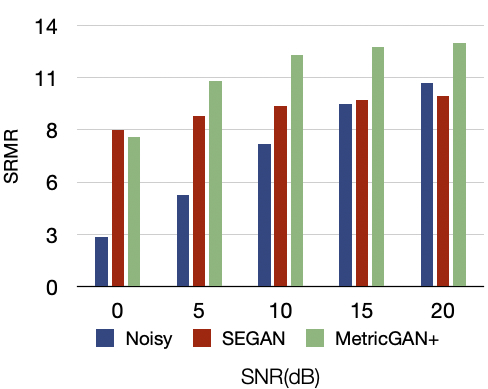}%
        \label{fig:cafeteria_srmr1}}
    \hfil
    \subfloat[Babble (SRMR)]{%
        \includegraphics[width=0.48\columnwidth]{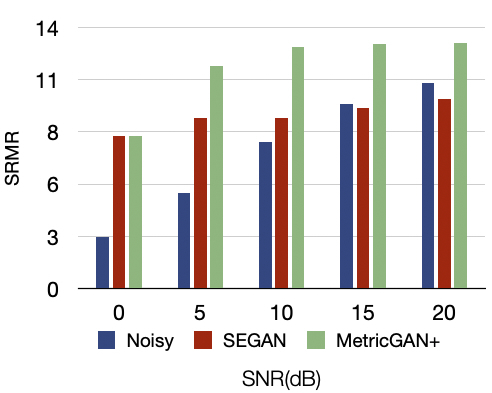}%
        \label{fig:babble_srmr1}}
    \caption{Perceptual quality performance of two enhancement algorithms based on PESQ and SRMR.}
    \label{fig:quality}
\end{figure}

\section{Experimental Results}
\label{sec:results}

In this section, we compare the performance of the two speech enhancement algorithms presented in section \ref{subsubsec:metricgan}. Figure \ref{fig:quality} shows the performance of the two algorithms in terms of PESQ and SRMR. MetricGAN+ provides the highest PESQ and SRMR scores for all noise level considered, while SEGAN led to lower scores for both measures. Although, both models are based on adversarial training, MetricGAN+ was developed based on its previous version which is optimized to increase the PESQ function \cite{fu2021metricgan+}. We assess the performance of these algorithms in terms of EER and t-DCF on the LA task. The proposed model achieves 1.4 and 0.04, respectively, for EER and t-DCF. Table \ref{tab:eer_tdcf_enhan}  shows the impact of background noise on the proposed classifier. Both adopted enhancement algorithms effectively mitigate the detrimental effects of noisy speecch. For cafeteria noises, EER performance improved for speech signals distorted by 0 dB, 5 dB, 10 dB, and 15 dB, with SEGAN outperforming MetericGAN+ at every instance. SEGAN is able to enhance our performance at 0 dB from 42.58 EER to as close as 14.03 EER, whereas MetricGAN+ can only enhance the performance from 42.58 to 40.40. The MetricGAN+ enhancement may be eliminating crucial audio cues necessary for spoof detection, as evidenced by Figure \ref{noisySNR}, which illustrates that the spectrogram for MetricGAN+ discards significant information in comparison to SEGAN. As the SNR levels rise, the performance of both enhancements becomes non-differential. At 15dB, our performance for SEGAN improves from 7.92 to 6.58 EER. As for MetricGAN+, our performance drops from 7.92 EER to 8.21 EER whereas for babble noises, EER performance improved for speech signals distorted by 0 dB, 5 dB, 10 dB, and 15 dB, with SEGAN outperforming MetericGAN+ at every instance. MetricGAN+ appears to have a larger EER at 0db, which could be due to the optimizer or the hyperparameters employed. SEGAN is able to enhance our performance at 0dB from 32.44 EER to as close as 15.21 EER, which is a significant increase in the improvement of the workload, whereas as noise levels rise, the performance of both enhancements becomes non-differential. At 15dB, our performance for SEGAN improves from 8.75 to 6.16 EER. As for MetricGAN+, our performance improves from 8.75 EER to 6.29 EER. At 20dB for both cafeteria and babble, it is evident that noisy levels perform far better than our augmentation strategies; this could be due to the fact that at 20dB there is less noise and a cleaner signal present.

\begin{table*}
\fontsize{10pt}{10pt}\selectfont
\caption{EER (\%) and t-DCF for SEGAN, MetricGAN+, and Noisy conditions across different noise levels}
\label{tab:eer_tdcf_enhan}
\centering
\scalebox{0.75}{
\begin{tabular}{c l c c c c c c c c c c}
    \toprule
    & & \multicolumn{2}{c}{\textbf{0 dB}} & \multicolumn{2}{c}{\textbf{5 dB}} & \multicolumn{2}{c}{\textbf{10 dB}} & \multicolumn{2}{c}{\textbf{15 dB}} & \multicolumn{2}{c}{\textbf{20 dB}}  \\  
    \cmidrule(lr){3-4} \cmidrule(lr){5-6} \cmidrule(lr){7-8} \cmidrule(lr){9-10} \cmidrule(lr){11-12} 
    & & \textbf{EER} & \textbf{t-DCF} & \textbf{EER} & \textbf{t-DCF} & \textbf{EER} & \textbf{t-DCF} & \textbf{EER} & \textbf{t-DCF} & \textbf{EER} & \textbf{t-DCF}\\ 
    \midrule
    \parbox[t]{2mm}{\multirow{3}{*}{\rotatebox[origin=c]{90}{Cafeteria}}} & SEGAN        & 14.03 & 0.33  & 11.91 & 0.29  & 8.99  & 0.22  & 6.58  & 0.15  & 5.89    & 0.16  \\  
    & MetricGAN+   & 40.40 & 0.90  & 23.20 & 0.65  & 13.47 & 0.42  & 8.21  & 0.25  & 5.47    & 0.17  \\  
    & Noisy        & 42.58 & 0.99  & 36.83 & 0.94  & 12.24 & 0.35  & 7.92  & 0.24  & 2.80    & 0.08  \\\\\\
    \parbox[t]{2mm}{\multirow{3}{*}{\rotatebox[origin=c]{90}{Babble}}} & SEGAN        & 15.21 & 0.40  & 10.41 & 0.26  & 7.56  & 0.20  & 6.16  & 0.17  & 5.70   & 0.16 \\  
    & MetricGAN+   & 42.80 & 0.94  & 24.17 & 0.67  & 11.23 & 0.34  & 6.29  & 0.34  & 3.60    & 0.09  \\  
    & Noisy        & 32.44 & 0.86  & 28.37 & 0.75  & 20.09 & 0.55  & 8.75  & 0.25  & 2.97    & 0.09  \\  
    \bottomrule
\end{tabular}}
\end{table*}

\begin{figure}[!t]
    \centering
    \subfloat[SEGAN (EER)]{%
        \includegraphics[width=0.48\columnwidth]{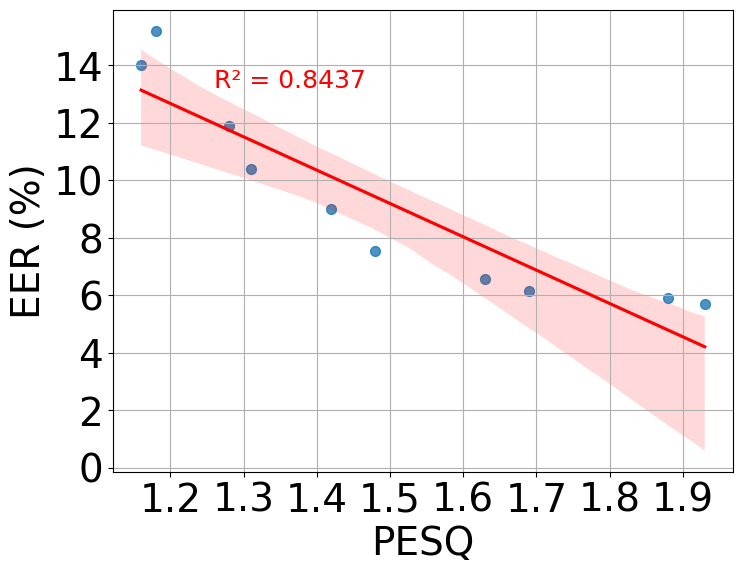}%
        \label{fig:seganeer}}
    \hfil
    \subfloat[MetricGAN+ (EER)]{%
        \includegraphics[width=0.48\columnwidth]{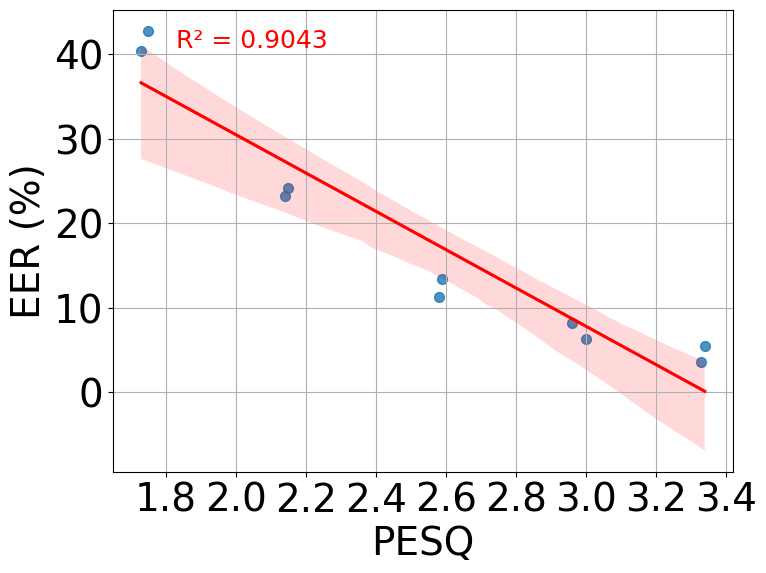}%
        \label{fig:meganeer}}
    \\[1ex]
    \subfloat[SEGAN (t-DCF)]{%
        \includegraphics[width=0.48\columnwidth]{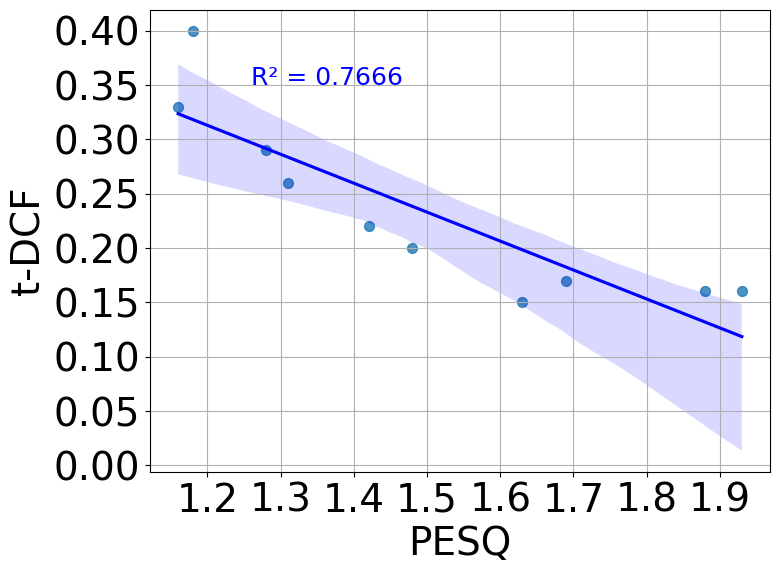}%
        \label{fig:segantdcf}}
    \hfil
    \subfloat[MetricGAN+ (t-DCF)]{%
        \includegraphics[width=0.48\columnwidth]{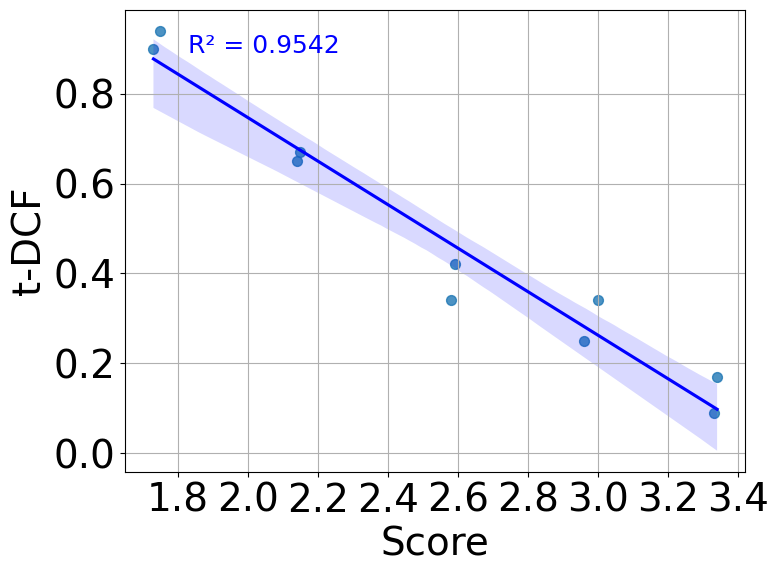}%
        \label{fig:megantdcf}}
    \\[1ex]
    \subfloat[SEGAN (EER, SRMR)]{%
        \includegraphics[width=0.48\columnwidth]{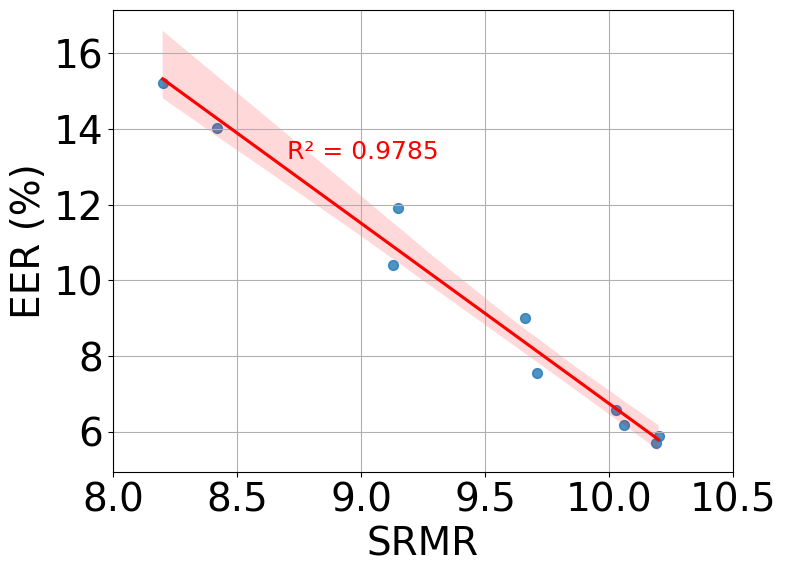}%
        \label{fig:seganeer_srmr}}
    \hfil
    \subfloat[MetricGAN+ (EER, SRMR)]{%
        \includegraphics[width=0.48\columnwidth]{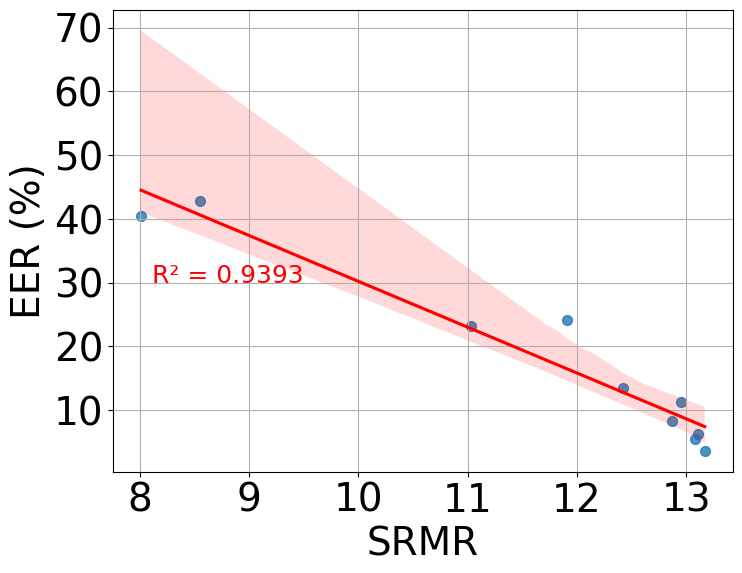}%
        \label{fig:meganeer_srmr}}
    \\[1ex]
    \subfloat[SEGAN (t-DCF, SRMR)]{%
        \includegraphics[width=0.48\columnwidth]{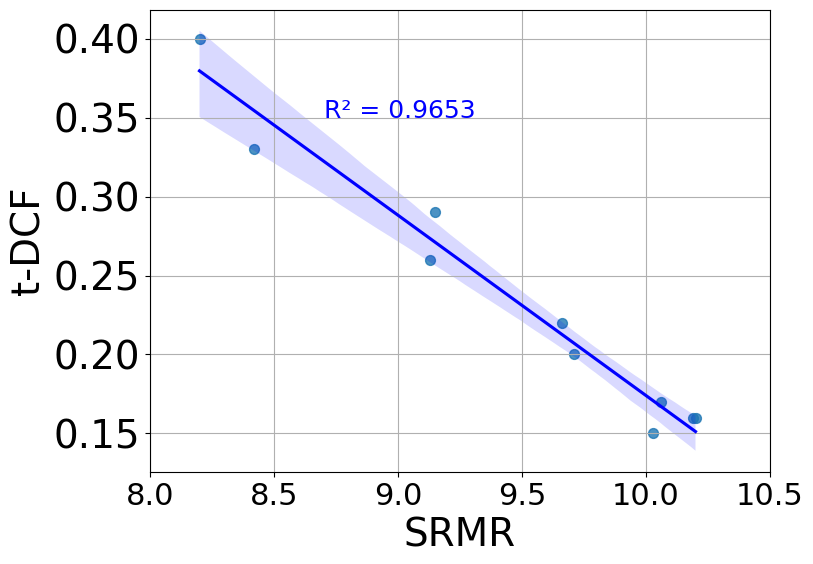}%
        \label{fig:segantdcf_srmr}}
    \hfil
    \subfloat[MetricGAN+ (t-DCF, SRMR)]{%
        \includegraphics[width=0.48\columnwidth]{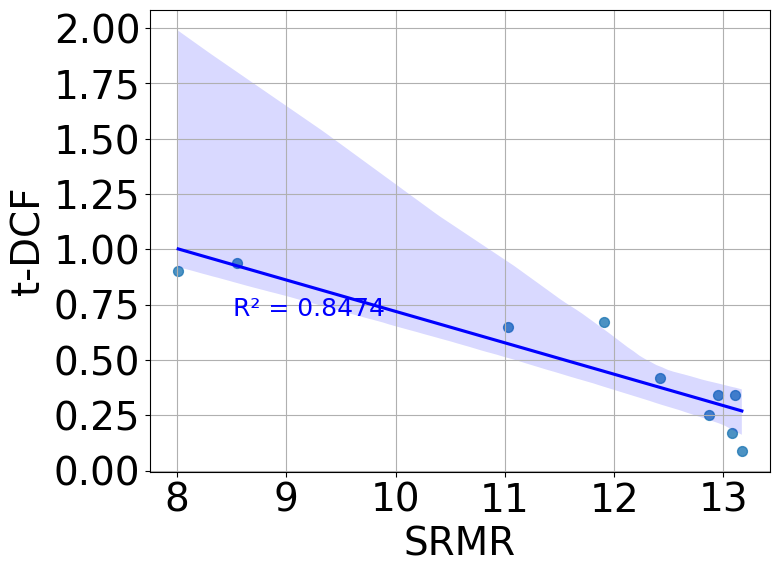}%
        \label{fig:megantdcf_srmr}}
    \caption{Correlation between PESQ/SRMR scores and EER, t-DCF for SEGAN and MetricGAN+. 
    (a–d) show correlations between PESQ vs. SEGAN/MetricGAN+ for EER and t-DCF, 
    (e–h) show correlations between SRMR vs. SEGAN/MetricGAN+ for EER and t-DCF.}
    \label{fig:corr}
\end{figure}

Figure \ref{fig:corr} shows the correlations between PESQ, SRMR measures and ASV system performance for both enhanced speech (SEGAN and MetricGAN+) and raw speech. Figure \ref{fig:corr} shows that MetricGAN+ has the highest correlation with PESQ measures (R2=0.90) for EER, while  SEGAN has the highest correlation with SRMR measures (R2=0.97) for EER, and MetricGAN+ has the strongest correlation with PESQ measures for t-DCF (R2=0.95) but SEGAN has the strongest correlation with SRMR measures for t-DCF (R2=0.96).
MetricGAN+ is specifically tailored for the quality aspect, whereas SEGAN performed well in the detection aspect. This reveals the need for a model that effectively integrates both aspects simultaneously.

\section{Conclusion}
\label{sec:conclusion}

In this paper, we evaluated the performance of an anti-spoofing system (AASIST) under varying noise conditions and investigated how speech enhancement affects audio deepfake detection. Two enhancement algorithms, SEGAN and MetricGAN+, were applied to mitigate the detrimental effects of noisy speech. As shown in our results, with noise present, our EER is as high as 42\%, with enhanced speech reducing the EER is to as low as 15\%. It is obvious that, despite the fact that our speech enhancement techniques are effective, SEGAN has a lower correlation than MetricGAN+ with quality measures such as PESQ and SRMR. Although SEGAN achieved lower scores for the quality measures, the processed speech by the algorithm achieved the lowest EER. MetricGAN+, on the other hand, provided the highest scores for both quality measures, but fell short in terms of achieving low EER. This may be attributed to the diverse datasets employed for pre-training, since MetricGAN+ emphasizes quality features, whereas SEGAN demonstrates superior audio generalization. Another hypothesis is that the enhancement of noisy data eliminates not just noise, but also crucial cues required for differentiating between authentic and spoofed audio. Our results reiterate that MetricGAN+ may discard some significant features in comparison to SEGAN. Both measure achieved high correlations with the performance of the LA task, with MetricGAN+ outperforming SEGAN.
Future work will evaluate the proposed approach on more recent datasets such as ASVspoof 5 to better reflect real-world conditions. Additionally, newer speech enhancement architectures, including Mamba-based models, will be investigated.
 

\bibliographystyle{IEEEtran}
\bibliography{refs25}

@string{icassp = "Proc. ICASSP"}

@string{interspeech = "Proc. Interspeech"}

@string{iwaenc = "Proc. IWAENC"}

@article{yamagishi2021asvspoof,
  author = {Junichi Yamagishi et al.},
  title = {ASVspoof 2021: Accelerating Progress in Spoofed and Deepfake Speech Detection},
  journal = {arXiv},
  year = {2021},
  eprint = {2109.00537},
  archivePrefix = {arXiv}
}

@article{kshirsagartask,
  title={Task-specific speech enhancement and data augmentation for improved multimodal emotion recognition under noisy conditions},
  author={Kshirsagar, Shruti and Pendyala, Anurag and Falk, Tiago H}
}

@article{wu2017asvspoof,
  title={ASVspoof: the automatic speaker verification spoofing and countermeasures challenge},
  author={Wu, Zhizheng and Yamagishi, Junichi and Kinnunen, Tomi and Hanil{\c{c}}i, Cemal and Sahidullah, Mohammed and Sizov, Aleksandr and Evans, Nicholas and Todisco, Massimiliano and Delgado, Hector},
  journal={IEEE Journal of Selected Topics in Signal Processing},
  volume={11},
  number={4},
  pages={588--604},
  year={2017},
  publisher={IEEE}
}

@article{pascual2017segan,
  title={SEGAN: Speech enhancement generative adversarial network},
  author={Pascual, Santiago and Bonafonte, Antonio and Serra, Joan},
  journal={arXiv preprint arXiv:1703.09452},
  year={2017}
}

@article{todisco2019asvspoof,
  title={ASVspoof 2019: Future horizons in spoofed and fake audio detection},
  author={Todisco, Massimiliano and Wang, Xin and Vestman, Ville and Sahidullah, Md and Delgado, H{\'e}ctor and Nautsch, Andreas and Yamagishi, Junichi and Evans, Nicholas and Kinnunen, Tomi and Lee, Kong Aik},
  journal={arXiv preprint arXiv:1904.05441},
  year={2019}
}

@inproceedings{yi2022add,
  title={Add 2022: the first audio deep synthesis detection challenge},
  author={Yi, Jiangyan and Fu, Ruibo and Tao, Jianhua and Nie, Shuai and Ma, Haoxin and Wang, Chenglong and Wang, Tao and Tian, Zhengkun and Bai, Ye and Fan, Cunhang and others},
  booktitle={ICASSP 2022-2022 IEEE International Conference on Acoustics, Speech and Signal Processing (ICASSP)},
  pages={9216--9220},
  year={2022},
  organization={IEEE}
}

@misc{speechbrain,
  title={{SpeechBrain}: A General-Purpose Speech Toolkit},
  author={Mirco Ravanelli and Titouan Parcollet and Peter Plantinga and Aku Rouhe and Samuele Cornell and Loren Lugosch and Cem Subakan and Nauman Dawalatabad and Abdelwahab Heba and Jianyuan Zhong and Ju-Chieh Chou and Sung-Lin Yeh and Szu-Wei Fu and Chien-Feng Liao and Elena Rastorgueva and François Grondin and William Aris and Hwidong Na and Yan Gao and Renato De Mori and Yoshua Bengio},
  year={2021},
  eprint={2106.04624},
  archivePrefix={arXiv},
  primaryClass={eess.AS},
  note={arXiv:2106.04624}
}

@article{fu2021metricgan+,
  title={Metricgan+: An improved version of metricgan for speech enhancement},
  author={Fu, Szu-Wei and Yu, Cheng and Hsieh, Tsun-An and Plantinga, Peter and Ravanelli, Mirco and Lu, Xugang and Tsao, Yu},
  journal={arXiv preprint arXiv:2104.03538},
  year={2021}
}

@inproceedings{jung2022aasist,
  title={Aasist: Audio anti-spoofing using integrated spectro-temporal graph attention networks},
  author={Jung, Jee-weon and Heo, Hee-Soo and Tak, Hemlata and Shim, Hye-jin and Chung, Joon Son and Lee, Bong-Jin and Yu, Ha-Jin and Evans, Nicholas},
  booktitle={ICASSP 2022-2022 IEEE international conference on acoustics, speech and signal processing (ICASSP)},
  pages={6367--6371},
  year={2022},
  organization={IEEE}
}

@article{de2024pesqetarian,
  title={The PESQetarian: On the Relevance of Goodhart's Law for Speech Enhancement},
  author={de Oliveira, Danilo and Welker, Simon and Richter, Julius and Gerkmann, Timo},
  journal={arXiv preprint arXiv:2406.03460},
  year={2024}
}

@article{avila2021automatic,
  title={Automatic speaker verification from affective speech using Gaussian mixture model based estimation of neutral speech characteristics},
  author={Avila, Anderson R and O’Shaughnessy, Douglas and Falk, Tiago H},
  journal={Speech Communication},
  volume={132},
  pages={21--31},
  year={2021},
  publisher={Elsevier}
}

@article{arif2021voice,
  title={Voice spoofing countermeasure for logical access attacks detection},
  author={Arif, Tuba and Javed, Ali and Alhameed, Mohammed and Jeribi, Fathe and Tahir, Ali},
  journal={IEEE Access},
  volume={9},
  pages={162857--162868},
  year={2021},
  publisher={IEEE}
}

@article{liu2023asvspoof,
  title={Asvspoof 2021: Towards spoofed and deepfake speech detection in the wild},
  author={Liu, Xuechen and Wang, Xin and Sahidullah, Md and Patino, Jose and Delgado, H{\'e}ctor and Kinnunen, Tomi and Todisco, Massimiliano and Yamagishi, Junichi and Evans, Nicholas and Nautsch, Andreas and others},
  journal={IEEE/ACM Transactions on Audio, Speech, and Language Processing},
  volume={31},
  pages={2507--2522},
  year={2023},
  publisher={IEEE}
}

@article{hanilci2016spoofing,
  title={Spoofing detection goes noisy: An analysis of synthetic speech detection in the presence of additive noise},
  author={Hanilci, Cemal and Kinnunen, Tomi and Sahidullah, Md and Sizov, Aleksandr},
  journal={Speech Communication},
  volume={85},
  pages={83--97},
  year={2016},
  publisher={Elsevier}
}

@inproceedings{chen2020generalization,
  title={Generalization of Audio Deepfake Detection.},
  author={Chen, Tianxiang and Kumar, Avrosh and Nagarsheth, Parav and Sivaraman, Ganesh and Khoury, Elie},
  booktitle={Odyssey},
  pages={132--137},
  year={2020}
}

@inproceedings{das2021data,
  title={Data augmentation with signal companding for detection of logical access attacks},
  author={Das, Rohan Kumar and Yang, Jichen and Li, Haizhou},
  booktitle={ICASSP 2021-2021 IEEE International Conference on Acoustics, Speech and Signal Processing (ICASSP)},
  pages={6349--6353},
  year={2021},
  organization={IEEE}
}

@inproceedings{avila2018investigating,
  title={Investigating Speech Enhancement and Perceptual Quality for Speech Emotion Recognition.},
  author={Avila, Anderson R and Alam, Md Jahangir and O'Shaughnessy, Douglas D and Falk, Tiago H},
  booktitle={Interspeech},
  pages={3663--3667},
  year={2018}
}

@inproceedings{santos2014improved,
  title={An improved non-intrusive intelligibility metric for noisy and reverberant speech},
  author={Santos, Jo{\~a}o F and Senoussaoui, Mohammed and Falk, Tiago H},
  booktitle={2014 14th International Workshop on Acoustic Signal Enhancement (IWAENC)},
  pages={55--59},
  year={2014},
  organization={IEEE}
}

@inproceedings{fu2019metricgan,
  title={Metricgan: Generative adversarial networks based black-box metric scores optimization for speech enhancement},
  author={Fu, Szu-Wei and Liao, Chien-Feng and Tsao, Yu and Lin, Shou-De},
  booktitle={International Conference on Machine Learning},
  pages={2031--2041},
  year={2019},
  organization={PmLR}
}

@inproceedings{tak2022rawboost,
  title={Rawboost: A raw data boosting and augmentation method applied to automatic speaker verification anti-spoofing},
  author={Tak, Hemlata and Kamble, Madhu and Patino, Jose and Todisco, Massimiliano and Evans, Nicholas},
  booktitle={ICASSP 2022-2022 IEEE International Conference on Acoustics, Speech and Signal Processing (ICASSP)},
  pages={6382--6386},
  year={2022},
  organization={IEEE}
}

@inproceedings{oyewale2024optimizing,
  title={Optimizing Voice Biometric Verification in Banking with Machine Learning for Speaker Identification},
  author={Oyewale, Oyebode Oluwatobi and Hossain, Md Delwar and Taenaka, Yuzo and Kadobayashi, Youki},
  booktitle={2024 IEEE 29th Asia Pacific Conference on Communications (APCC)},
  pages={377--384},
  year={2024},
  organization={IEEE}
}

@article{narang2024artificial,
  title={Artificial intelligence in banking and finance},
  author={Narang, Ashima and Vashisht, Priyanka and Bajaj, Shalini Bhaskar},
  journal={International Journal of Innovative Research in Computer Science and Technology},
  volume={12},
  number={2},
  pages={130--134},
  year={2024}
}

@inproceedings{rix2001perceptual,
  title={Perceptual evaluation of speech quality (PESQ)-a new method for speech quality assessment of telephone networks and codecs},
  author={Rix, Antony W and Beerends, John G and Hollier, Michael P and Hekstra, Andries P},
  booktitle={2001 IEEE international conference on acoustics, speech, and signal processing. Proceedings (Cat. No. 01CH37221)},
  volume={2},
  pages={749--752},
  year={2001},
  organization={IEEE}
}

\end{document}